\documentclass[10pt,a4paper,twoside]{article}
\usepackage{epsfig}
\usepackage{baltlat6}
\long\def\symbolfootnote[#1]#2{\begingroup%
\def\thefootnote{\fnsymbol{footnote}}\footnote[#1]{#2}\endgroup}
 
\begin{document}
\ \
\vspace{0.5mm}
\setcounter{page}{1}
\vspace{8mm}
 
\titlehead{BAASP, 2012}
 
\titleb{ON THE VARIABILITY OF TOTAL SOLAR IRRADIANCE}
 
\begin{authorl}
\authorb{J.~Pelt}{} and
\authorb{O.~K\"arner}{}
\end{authorl}
 
\begin{addressl}
\addressb{}{Tartu Observatory, T\~{o}ravere, 61602, Estonia}
\end{addressl}
 
\submitb{Received: 2012 x x; accepted: xxxx x x}
 
Knowing the variability of total solar irradiance  (TSI) at the top of the  atmosphere  is crucial 
for specifying solar influence to climate  variability. Satellite measured TSI data are available 
only for the last  32 years. 
But there is an opportunity to estimate approximate daily TSI  values on the basis of the 
observed variability of solar activity. 
A common approach is based using data for Wolf sunspot numbers. 
Series of daily data from the DAVOS based TSI starting from 1978  have been compared with 
daily values of Wolf numbers and with the corresponding estimates for daily summary sunspot area. 
Our goal is to use established correlations to hindcast earlier TSI values. 
The available solar activity data enabled us to reproduce rough TSI estimates back to the 1870's.
 
 
 
\resthead{Towards the automatic estimation of time delays}
{J.~Pelt and O. K\"arner}

\sectionb{1}{INTRODUCTION}
 
Modern climatological thinking is based on the dominating role of the anthropogenic global warming theory
(Solomon et al. 2007).
It is supported by continuous increase of  the concentration of carbon dioxide in the Earth's atmosphere.
To estimate validity of that conception it is necessary to know the temporal variability of 
other forcing factors, among them the total solar irradiance (TSI) at the top of the atmosphere.
Direct satellite measurements on that flux have been carried out only since 1978. 
But this flux density depends on solar magnetic activity describable by Wolf numbers or other indicators. Knowing 
the relationship between solar activity and TSI enables us to hindcast the TSI series more than 100 years 
back on the basis of observations of sunspots. This enables one to estimate changes in solar forcing in
comparison with that caused by the increase of CO$_2$ concentration in the Earth's atmosphere.    
 
Solar activity is often characterized by means of sunspot numbers (or Wolf numbers)
The corresponding quantitative characteristic $R_w$ is computed as $R_w = k(10g + s)$, where
$s$  is number of distinct spots (independent on their size); $G$ is number of spot groups and $k$ is 
a coefficient to reconcile data with different observation places and telescopes. The other activity 
indicator used in the current study is a daily sunspot area $A_S$, also available from historical 
records.  
 
A time series for daily $R_w$ has been observed from 1870. 
Satellite based TSI observations started in 1978. Two types of observations have been carried out 
simultaneously during the last three decades. This enables us to to empirically estimate 
statistical relationship between daily values of $R_w$, $A_S$ and TSI. 
Physical basis of that relationship has been explained recently (e.g. Foukal et al. 2006):
{\em Brightening of the Sun with increasing sunspot number is explained by the existence of bright 
magnetic areas called faculae. Each such dark or bright structure contributes to a TSI variation
equal to the product of its projected area and its photometric contrast relative to the adjacent,
undisturbed photometric disc.} 
 
Much higher contrast of the faculae causes increase of TSI.
Thus an examination of statistical relationship between daily values of $R_w$, $A_S$ and TSI is
justified. 
 
Several attempts to hindcast previous TSI values have been performed using different
time intervals. Diagnosed associations between sources of contemporary irradiance variability and
appropriate solar activity proxies that extend over longer time spans permit the 
irradiance reconstructions for some earlier period 
(e.g. Lean 1997, Wenzler et al. 2006, Balmaceda et al. 2007, Steinhilber et al. 2009, Krivova et al. 2010, Shapiro et al. 2011). 
New reconstructions 
of spectral irradiance are developed for time base starting from 1600 with absolute scales traceable to 
space based observations. For instance an updated solar irradiance reconstruction (Lean 2000) on annual resolution 
is available online (Lean 2004).
 
In this paper we compare two methods to hindcast TSI values for earlier than observed dates. The first method
is rough and is based on regression analysis of TSI {\it versus} $R_w$. The second method is a new and uses 
daily sunspot areas and corresponding TSI values to build empirical model  for relation between the two observed series.
The model is based on the observation that solar magnetic activity {\em correlates} with low-frequency flow
of the TSI and at the same time {\em anti-correlates} with high frequency features. Our empirical model
contains free parameters but values for them can be obtained from fitting procedure itself. Thereby our
method is essentially parameter free and differs from traditional methods where modelling is based on
heavily parametrized physical models (see for instance Krivova et al. 2007).     
 
\sectionb{2}{DATA} 
 
For general overview of the available TSI data and processing methods see e.g. Domingo et al. 2009.  
The original TSI satellite observations are compiled into homogeneus data sets, often called composites
(for recent overviews and compiling methods see Ball et al. 2012 and Domingo et al. 2009).
Results for daily TSI in Wm$^{-2}$ for the time interval 1978.88-2012.02 have been downloaded from the website 
belonging to Davos Physical-Meteorological Observatory (PMOD)\footnote
{\tt ftp://ftp.pmodwrc.ch/pub/data/irradiance/composite}
(see Fr\"ohlich and Lean 1998 for details
and for updates Fr\"ohlich 2006). All together there is 11428 adjusted measurements in composite data.  
The observation archives from Belgian Royal Observatory  were used to 
download daily sunspot numbers\footnote{\tt http://www.sidc.be/DATA/dayssn\_import.dat} 
from 1818.02 to 2012.16 - all togeteher 67670 estimates.
Sunspot daily areas were downloaded from NASA/Marshall Space Flight Center Solar Physics   
web pages\footnote{\tt http://solarscience.msfc.nasa.gov/greenwch/daily\_area.txt}. This data set contains
50212 measurements from 1874.35 to 2012.16.
 
\sectionb{3}{PRELIMINARY ANALYSIS}
 
\begin{figure}
\begin{center}
\includegraphics[height=55mm]{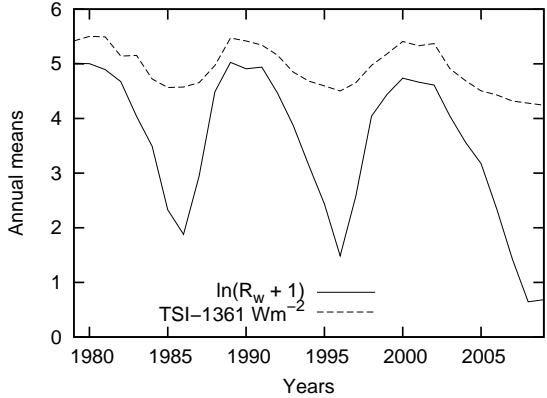}
\caption{Comparison of the annual mean values for ln(R$_w$ +1.) TSI anomalies i.e. ($TSI - 1361$) in  Wm$^{-2}$
during the last 30 years.} \label{wo-tsi-1}
\end{center}
\end{figure}

Annual mean values over the period from 16.11.1978 to 20.09.2010 have been computed for the comparison. 
The results are shown in Figure  ~\ref{wo-tsi-1}.
In order to convert the draft better comparable, the $R_w$ values are presented via logarithm i.e.
ln(R$_w $+1.) and TSI data as difference from the constant 1361 Wm$^{-2}$.
 
Three important details can be found:
\begin{enumerate}
\item The both variables change in the same phase.
\item There is a remarkable difference in the length of the low activity periods.
\item The difference of TSI values between high and low activity seasons is in between $0.5-1$ Wm$^{-2}$.
\end{enumerate}
The website  {\tt gaw.kishou.go.jp/wdcgg.html} for the World Database for Greenhouse Gases  (WDCGG)
states that the concentration of CO$_2$ grows at constant rate 1.6 ppm/year.  Radiative transfer calculations 
(e.g. Schwartz 2007) have shown that the direct influence of doubling the concentration
(i.e. from 320 ppm to 640 ppm) to the outgoing longwave radiation (OLR) flux density is about 4 Wm$^{-2}$.
Provided that the growth rate does not change, it will take about $320/1.6 = 200$ years
until the concentration doubles. This means that the annual part in the OLR flux growth is
about $4/200 = 0.02$  Wm$^{-2}$. 
This value is significantly lower than the difference between TSI between high and low solar 
activity. The variability of TSI can in a such way dominate. However, its influence to the Earth climate remains
dependent on cumulative feedback in the climate system.

\sectionb{4}{APPROXIMATE DEPENDENCE of TSI on $R_w$}
 
The existing dataset enables us to display approximate dependence of TSI on $R_w$.
First, it is interesting to select spot free days.
During the solar minimum of 2008, the value of TSI was more than 0.2 Wm$^{-2}$ lower than during the 
previous minimum in 1996, indicating for the first time a directly observed long-term change (Fr\"ohlich 2009).
Direct comparison shows that the same difference occurs between the mean values for 
sunspot free days for the periods 1978-1999 and  2000-2010, respectively. 
Thus, for more accurate restoration we compare the daily $R_w$ and TSI values over the period
1978-2003 only. 
 
\begin{figure}
\begin{center}
\includegraphics[height=55mm]{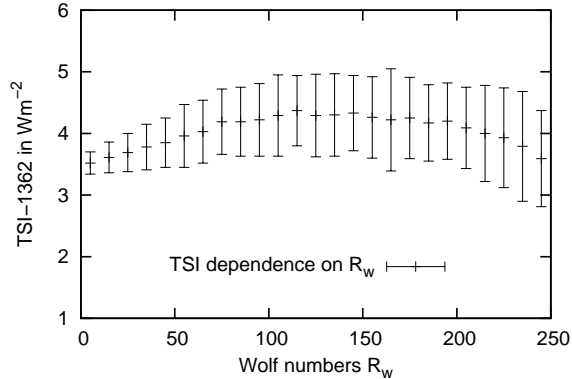}
\caption{Range of daily variability for $TSI - 1362$ depending on different level of $R_w$. The days corresponding to $R_w =0$ are excluded} \label{wo-tsi-2}
\end{center}
\end{figure}

Empirical dependence of TSI on $R_w >0$ is simple to figure out. In this case the region of variation (1 - 250) for $R_w$ is partioned to
10 unit bins.  As  a result 25 bins are selected. Values of daily TSI (on the basis of PMOD composite) 
for each bin have been collected and
their range is shown in Figure  ~\ref{wo-tsi-2} with error bars. The mean TSI values of bins are 
marked by straight lines.
 
Figure ~\ref{wo-tsi-2} shows that the dependence of TSI on $R_w$ 
is similar to the the arc produced by Solanki and Fligge (1999) and 
Hempelmann and Weber (2011). The latter study has been produced 
using ACRIM based TSI series. This is expected because both
TSI series have been composed from measurements of the same satellites.
 
Maximal values of  TSI correspond to $R_w$ from the interval $100 < R_w <150$. It appears to be somewhat more elaborate than
a linear expectation on proportionality between spots and faculi could produce. 
Thus, the oscillation of $R_w$ in between 0 and 250 causes an oscillation of TSI around somewhat higher 
level than that corresponding to TSI from a spotless Sun.  Anyway, the increase of TSI as $R_w$ is growing will be bounded. 
Approximate relationship between TSI and $R_w$ shown in Figure  ~\ref{wo-tsi-2} enables us to estimate 
TSI variability caused by the variations of solar activity.
In the current case a restoration is carried out using the mean TSI values for each 
$R_w$ interval presented in Figure ~\ref{wo-tsi-2} . The results will be presented together with those for
the other method in Figure ~\ref{wo-tsi-3}. 
 
\sectionb{6}{REFINED ANALYSIS USING $A_S$}
 
Rough analysis done so far can be refined significantly. It is possible to build, based on various physical assumptions, so called
model or proxy curves (see e.g. Krivova et al. 2007) and use them to hindcast daily TSI values. This 
kind of methods demand
complicated physical modelling for different magnetic structures on the surface of the Sun (spots, penumbrae, flares and network). In this paper we
try to hindcast using another parameter free scheme.
In the following a more detailed dataset of sunspot areas $A_S$ will be used instead of sunspot numbers. 
 
\begin{figure}
\begin{center}
\includegraphics[height=45mm]{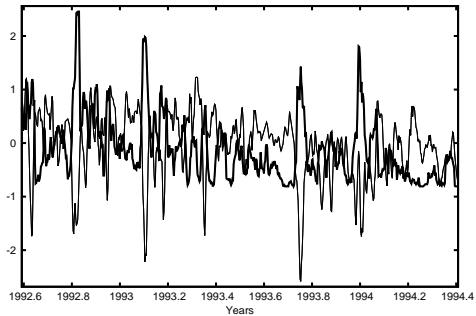}
\caption{Comparison of the normalised TSI (thin line) and sunspot area values (thick line). The smaller details in the curves tend to anti-correlate.} \label{compa-corr}
\end{center}
\end{figure}

It is well known that overall (low frequency) TSI flow quite well correlates with solar magnetic activity. 
But if to look at Figure~\ref{compa-corr} then it can be easily seen that
the smaller details anti-correlate. 
This can be easily explained referring to darker sunspots moving on the surface of the Sun. 
From the point of view of the time series analysis such combination of varying correlation 
introduces extra complications.
We need to model separately low frequency and high frequency parts of the input curves because 
of that difference.   
 
\subsectionb{6.1}{Envelopes}
 
Typically low frequency (smooth) components of the time series are obtained with different averaging and smoothing algorithms. 
In this work we used so called LOWESS (locally weighted scatterplot smoothing, see Cleveland 1979) method with linear downweighting to
compute low frequency components. This method is constrained with only single parameter - a smoothing window width $W$. By changing this parameter
it is possible to obtain curves with different degrees of smoothness.
 
The activity records of the Sun tend to be one sided. For instance sunspot areas are positive or zero. This forces us to consider additional
time series component types beside the trivial smoothed or averaged data. 
The method to build such an one sided smooth curves has been employed earlier for processing of spectral data 
where the absorption or emission lines are superposed on a smooth continuum (e.g. Pelt 1990). This is very similar to the situation
with sunspot blocking. 
 
\begin{figure}
\begin{center}
\includegraphics[height=55mm]{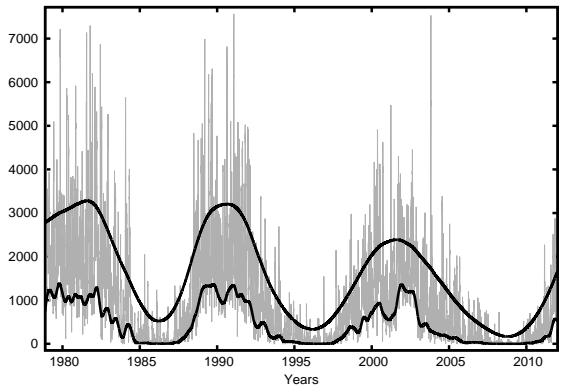}
\caption{Original sunspot area data in millionth parts of solar surface, 
upper envelope $E^{U}(3.0,4,t)$ and lower envelope $E^{L}(0.4,9,t)$ for the sunspot areas. 
In this time interval we can use the enevelopes as fit components to model TSI.} \label{envelopes}
\end{center}
\end{figure}

In our method the smoothing and one side clipping is performed
iteratively to obtain a smoothed component which envelopes the input time series. In the very end of the iterations the smooth envelope goes through the local
minima or maxima. We do not need perform all the iteration but can stop earlier, say at iteration $K$. Taking into account that smoothing operation itself depends
on the width parameter $W$ we can formally build from input time series 
two sets of envelopes: $E^{U}(W^{U},K^{U},t)$ and $E^{L}(W^{L},K^{L},t)$. The first set describes upper part
of the data and the second set lower part. An illustration of the envelopes is provided on Figure~\ref{envelopes}.      
 
\subsectionb{6.2}{Regression}
 
\begin{figure}
\begin{center}
\includegraphics[height=55mm]{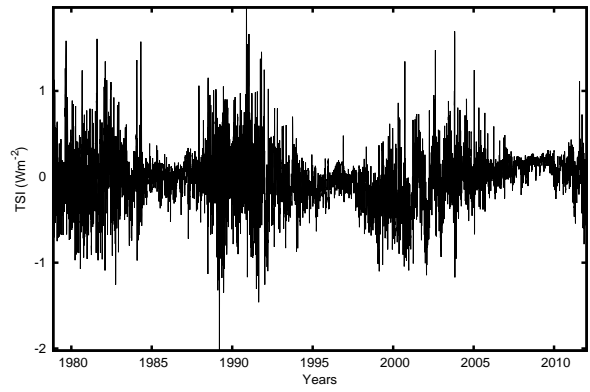}
\caption{Comparison of the observed  {\it versus} restored TSI values} \label{difference}
\end{center}
\end{figure}

The least squares distance $D^2$ between four plus four parameter model buildt from sunspot area data ($S(t)$) and TSI ($I(t)$):
\begin{equation}
   D^2=\big (a_0+a_1 E^{L}(W^{L},K^{L},t)+a_2 E^{U}(W^{U},K^{U},t)+a_3 S(t) - I(t) \big )^2
\end{equation} 
must be minimized to obtain proper estimates for the regression coefficents $a_0,a_1,a_2,a_3$ and also for smoothing parameters $W^{U}$,$K^{U}$,$W^{L}$ and $K^{L}$.
Combining linear least squares estimation with grid search for nonlinear parameters allows us to obtain final solution. The results for our particular case
(daily values of PMOD composite for TSI {\it versus} records of sunspot areas) can be illustrated in different ways. First we can plot the difference 
between actual TSI and computed from sunspot areas TSI
(see Figure~\ref{difference}). 
This plot can be compared with similar figure 4c from Krivova et al. (2007) where overall scatter of the differences between observed and predicted curves is somewhat
higher.
 
\begin{figure}
\begin{center}
\includegraphics[height=55mm]{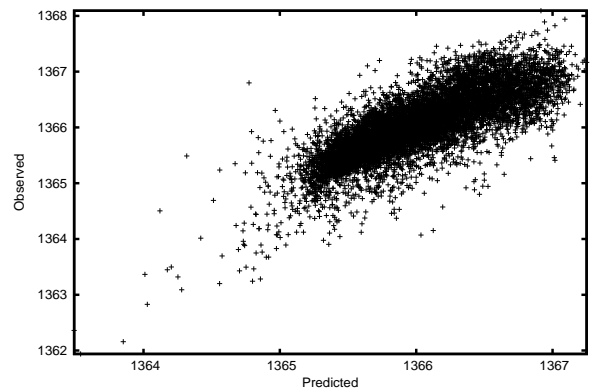}
\caption{Scatter plot of the observed  {\it versus} restored daily TSI values.} \label{cross}
\end{center}
\end{figure}

The cross plot between the original and restored TSI values is presented on Figure~\ref{cross}. There is still considerable scatter but the correlation level between
two series is quite high ($r=0.826$). 
As a matter of fact this value is comparable to the scatter between different TSI composites (correlation between ACRIM and PMOD composites is at level $r=0.834$). 
 
\begin{figure}
\begin{center}
\includegraphics[height=55mm]{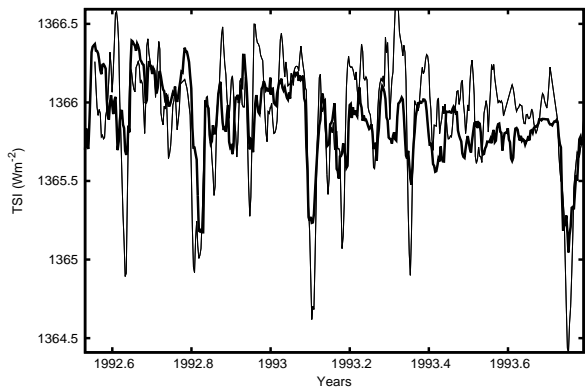}
\caption{Small details and larger features of the estimated and original curve are now correlated.} \label{compa2}
\end{center}
\end{figure}

\begin{figure}
\begin{center}
\includegraphics[height=55mm]{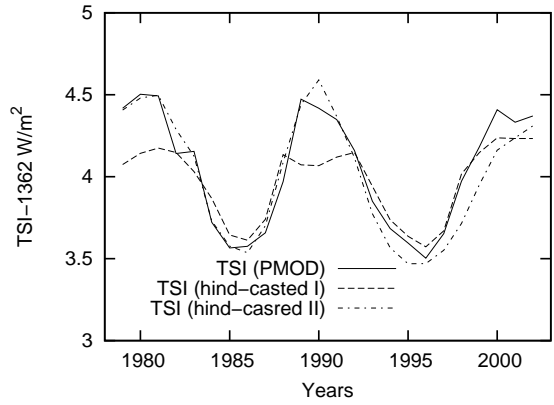}
\caption{Comparison of the measured annual mean $TSI_{PMOD}$  values with the annual
mean values computed using the two hindcasting schemes, respectively.} \label{wo-tsi-3}
\end{center}
\end{figure}

In our earlier observation we saw that small details in the flow of the sunspot area data are anti-correlated with small details of the TSI flow. 
However, the weighted combination of sunspot area data and its envelopes shows much better correlation. 
This is a result of estimated $a_3$ value being less than zero. Positive correlation of the large scale features comes from contributions of envelope components
$E^{L}(W^{L},K^{L},t)$ and $E^{U}(W^{U},K^{U},t)$. The situation is precisely illustrated on Figure~\ref{compa2}.
 
Finally, it is useful to illustrate the results of restoration of TSI over our testing interval. The illustration is shown 
in Figure~\ref{wo-tsi-3} in annual means of (TSI-1362) Wm$^{-2}$.
 
Figure ~\ref{wo-tsi-3} shows that the rough restoring operation has varying accuracy. For the regions of low solar activity
its behavior is acceptable. But for the regions with high solar activity it gives a significant bias.
This means that it is not applicable to use the scheme to hindcast previous TSI values. 
The second method (even after averaging) gives much better accordance with the original values over that interval. 
 
\begin{figure}
\begin{center}
\includegraphics[height=55mm]{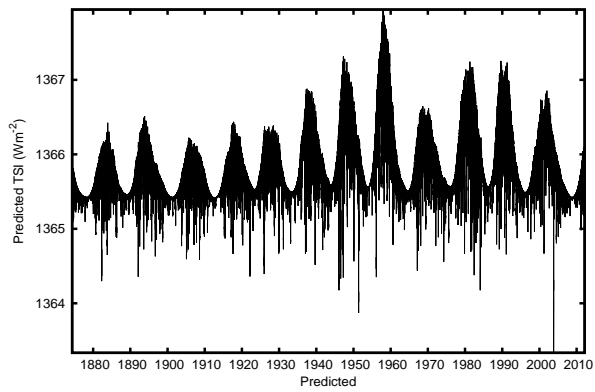}
\caption{The TSI values for previous years computed from the daily sunspot areas.} \label{predicted}
\end{center}
\end{figure}

The hindcasted daily values of TSI for the whole time interval with available $A_S$ data is presented at
Figure~\ref{predicted}. This series can be used in modelling programs or compared with temperature curves.
 
\sectionb{7}{CONCLUSIONS}
 
Hindcasting the previous TSI data is important in order to improve our understanding of climate variability.
In this paper we used two methods to hindcast past TSI values using indicators of the solar magnetic activity
as a base. The rough and more traditional regression method was supplanted by the refined method which uses
daily values of the sunspot areas. The refined method is essentially parameter free. This is important because the
traditional hindcast methods are based on a large set of parametric physical models of the solar features.
Here we demonstrate that the low frequency behaviour of the TSI is controlled by the low frequency components
of the magnetic activity. However, this dependence is not straight forward but somewhat convoluted.
Similar result was obtained by using signal theory methods of empirical data modelling (Preminger 2010). The obvious
advantage of proposed here method is its conceptual transparency - upper and lower envelopes are easily
grasped and visualized. We plan to use the new method in different contexts. 
     
Variability of the hindcasted  TSI values shows that the interannual variability of TSI has been nearly of the same 
range as that during the last three decades. 
Previous studies have shown a remarkable variability of the solar cycle length to influence the 
long-term variability in the climate system (e.g. Lassen and Friis-Christensen 1995). 
The described empirical relationship between TSI and solar activity enables one to examine 
such an influence in more detail.
 
\References
 
\refb Ball W.T, Unruh Y.C., Krivova N.A. et al. (2012).
Reconstruction of total solar irradiance 1974 - 2009. 
{\it arXiv:1202.3354v1},1-15. 
 
\refb Balmaceda L., Krivova N.A., Solanki S.K. (2007).
Reconstruction of solar irradiance using the Group sunspot number,
{\it Adv. Space Res., 40}, 986–989.
 
\refb Barnhart B.L., Eichinger W.E. (2011).
Empirical Mode Decomposition applied to solar irradiance,
global temperature,sunspot number,and $CO^2$ concentration data, 
{\it Journ. Atmos. Solar-Terr. Phys, 73}, 1771-1779.

\refb Cleveland W.S. (1979).
Robust Locally Weighted Regression and Smoothing Scatterplots 
{\it J. Amer. Stat. Assoc., 74}, 829–836.
 
\refb Domingo V., Ermolli I., Fox P. et al. (2009). 
Solar Surface Magnetism and Irradiance on Time Scales
from Days to the 11-Year Cycle, 
{\it Space Sci. Rev.,145},337-380. 
 
\refb Foukal P., Fr\"olich C., Spruit H., Wigley T.M.L., (2006). 
Variations in solar luminosity and their
effect on the Earth's climate. 
{\it Nature, 443}, 161-166. 
 
\refb Fr\"ohlich C. (2009).
Evidence of a long-termtrend in total solar irradiance. 
{\it Astron. and Astrophy.,
501},L27-L30. DOI: 10.1051/0004-6361/200912318.
 
\refb  Fr\"ohlich C., Lean J. (1998).
The Suns Total Irradiance: Cycles and
Trends in the past two decades and associated Climate Change Uncertainties.
{\it Geophys.Res.Lett., 25}, 4377-4380. 
 
\refb Fr\"ohlich, C., (2006).
Solar irradiance variability since 1978: Revision of the PMOD composite during solar cycle
21, 
{\it Space Sci. Res., 125}, 53–65.

\refb  Hempelman A., W. Weber (2011).
Correlation between the sunspot number, the total solar irradiance, and the terrestrial
insolation. 
{\it Solar Phys., 277}, 417-430,
DOI 10.1007/s11207-011-9905-4.

\refb  Krivova N.A., Balmaceda L., Solanki S.K. (2007).
Reconstruction of solar total irradiance since 1700
from the surface magnetic flux, 
{\it Astron. and Astroph., 467}, 335-346.
 
\refb  Krivova N.A., Vieira E.A., Solanki S.K. (2011). 
Reconstruction of solar spectral irradiance since
the Maunder minimum, 
{\it J. Geophys Res., 115}, A12112,1-11.  
 
\refb Lassen K., Friis-Christensen E. (1995).
Variability of the solar cycle length during the past five centuries and the 
apparent association with terrestrial climate. 
{\it J. Atmos. Solar.-Terr. Phys., 57}, 835-845.  
 
\refb Lean J. (1997).
The Sun's variable radiation and its relevance for earth. 
{\it Rev. Astron. Astrophys., 35}, 33-67.
 
\refb  Lean J. (2000).
Evolution of the Sun's Spectral Irradiance Since the Maunder Minimum.
{\it Geophys. Res. Lett., 27},2425-2428.

\refb  Lean J. (2004).
Irradiance Reconstruction. 
{\it IGBP PAGES/World Data Center
for Paleoclimatology. Data Contribution Series No 2004-035}. NOAA/NGDC Paleoclimatology
Program, Boulder, CO, USA. 
 
\refb  Pelt J. (1990).
A Method of Continuum Estimation in Spectral Classification Experiments, 
{\it Bulletin d'Information du Centre de Donnees Stellaires},  38, 95-107.
 
\refb  Preminger D., Nandy D., Chapman G., Martens P.C.H (2010).
Empirical Modeling of Radiative versus Magnetic Flux
for the Sun-as-a-Star. 
{\it Solar Physics, 264}, 13-30.
 
\refb Schwartz S.E. (2007).
Heat capacity, time constant, and sensitivity of Earth's climate system. 
{\it J. Geophys. Res., 112, D24S05},1-12.  
 
\refb Solomon S., Qin S., Manning M. et al. eds. (2007). {\it Climate Change, 
The Physical Science basis}, Cambridge University Press.
 
\refb Shapiro A.I., Schmutz W., Rozanov E., Schoell M., Haberreiter M., Shapiro A.V., Nyeki S. (2012).
A new approach to the long-term reconstruction of the solar
irradiance leads to large historical solar forcing, 
{\it Astron. and Astrophys., 529}, A67, 1-8.
 
\refb Solanki S.K., M. Fligge (1999).
A reconstruction of total solar irradiance
since 1700. {\it Geophys. Res. Lett., 26}, 2465-2468. 
 
\refb Steinhilber F., Beer J.,  Fr\"ohlich C. (2009).
Total solar irradiance during the Holocene,
{\it Geophys. Res. Lett., 36}, L19704,1-5. 
 
\refb Wenzler T., Solanki S.K., Krivova N.A., Fr\"ohlich C. (2006).
Reconstruction of solar irradiance variations in cycles 21–-23 based on surface magnetic fields
{\it Astron. and Astrophys. 460}, 583
-595
 
\end{document}